\documentclass[twocolumn,twocolappendix]{aastex701}

\usepackage{amsmath}
\usepackage{MnSymbol}
\usepackage{rotating}
\usepackage{array}
\usepackage{eucal}
\usepackage{color}
\usepackage{xcolor}
\usepackage{graphicx}
\usepackage{mathrsfs}
\usepackage{color,soul} 
\usepackage{subfiles}
\usepackage{hyperref}
\usepackage[T1]{fontenc}

\newcommand{\mat}[3]{\displaystyle\langle #1 | #2 | #3 \rangle}

\newcommand{\threeJ}[6]{\begin{pmatrix}#1 & #2 &#3 \\ #4& #5 & #6\end{pmatrix}}
\newcommand{\sixJ}[6]{\begin{Bmatrix}#1 & #2 &#3 \\ #4& #5 & #6\end{Bmatrix}}
\newcommand{\nineJ}[9]{\begin{Bmatrix}#1 & #2 &#3 \\ #4& #5 & #6\\ #7& #8 & #9\end{Bmatrix}}
\newcommand{\X} {{\it Xenomorph}}
\newcommand{\B} {{\sc balrog}}

\newcommand{\EB} {\mathcal{H}}
\newcommand{\Av}[1]{\left\langle #1\right\rangle}

\newcommand{\E}{\psi}

\newcommand{\LT}{\mathfrak{L}}
\newcommand{\ST}{\mathfrak{S}}
\newcommand{\JT}{\mathfrak{J}}

\shorttitle{Line Broadening in Stellar Flares}
\shortauthors{Gomez et al.}

\begin{document}


\title{Enigmatic Line Broadening During Solar Flares: Magnetic Field Broadening?}

\correspondingauthor{T. A. Gomez}
\email{thomas.gomez@colorado.edu}

\author[0000-0001-8748-5466]{Thomas A. Gomez}
\email{thomas.gomez@colorado.edu}
\altaffiliation{George Ellery Hale Fellow}
\affiliation{Department of Nuclear and Radiation Sciences, University of Michigan, Ann Arbor, 48109, USA}
\affiliation{Department of Astrophysical and Planetary Sciences, University of Colorado, Boulder, CO 80305, USA}
\affiliation{Laboratory for Atmospheric and Space Physics, University of Colorado Boulder, Boulder, CO 80303, USA}
\affiliation{National Solar Observatory, University of Colorado Boulder, Boulder, CO 80303, USA}
\affiliation{Department of Astronomy, University of Texas at Austin, 
 Austin, TX 78712, USA}

\author[0000-0001-7458-1176]{Adam F. Kowalski}
\email{adam.f.kowalski@colorado.edu}
\affiliation{Department of Astrophysical and Planetary Sciences, University of Colorado, Boulder, CO 80305, USA}
\affiliation{Laboratory for Atmospheric and Space Physics, University of Colorado Boulder, Boulder, CO 80303, USA}
\affiliation{National Solar Observatory, University of Colorado Boulder, Boulder, CO 80303, USA}

\author[0000-0002-3229-1848]{Cole Tamburri}
\email{cole.tamburri@colorado.edu}
\affiliation{Department of Astrophysical and Planetary Sciences, University of Colorado, Boulder, CO 80305, USA}
\affiliation{Laboratory for Atmospheric and Space Physics, University of Colorado Boulder, Boulder, CO 80303, USA}
\affiliation{National Solar Observatory, University of Colorado Boulder, Boulder, CO 80303, USA}

 \author[0000-0001-5316-914X]{Graham S. Kerr}
 \email{graham.kerr.2@glasgow.ac.uk}
  \affiliation{SUPA School of Physics and Astronomy, University of Glasgow, Glasgow, G12 8QQ, UK}
 \affiliation{Department of Physics, Catholic University of America, 620 Michigan Ave Northeast, Washington D.C. 20064, USA}

 \author[0000-0003-4052-2746]{Jackson R. White}
 \email{jacksonwhite@utexas.edu}
 \affiliation{Department of Astronomy, University of Texas at Austin, 
  Austin, TX 78712, USA}
 \affiliation{X Computational Physics Division, Los Alamos National Laboratory, 
  Los Alamos, NM 87545, USA}

\date{\today}






\begin{abstract}
The origin of the extreme broadening observed in chromospheric metal lines during solar and stellar flares—particularly Mg~{\sc ii} h\&k and Ca~{\sc ii} H\&K—remains poorly understood.
These lines often display Lorentzian-like wings whose widths exceed standard Stark broadening predictions by factors of $\sim 30$, with no known collisional mechanism capable of producing such enhancements.
We posit that magnetic fields are responsible for this additional broadening due to the increase of magnetic activity during flares.
A magnetic-field distribution of the form $P(B)\propto B^{-3}$ reproduces the observed Mg {\sc ii} profile wings while leaving H {\sc i} Balmer lines and optically thin transitions largely unaffected.
{To explain the broadening using magnetic fields, the high $B$ tail can extend up to $10^6$G with extremely low probabilities where the filling factors are $\lesssim 10^{-6}$.}
We propose that observations of flares using spectropolarimetry can verify whether the anomalous broadening is from magnetic structures in flare ribbons.
\end{abstract}


\section{Introduction} 

Solar and stellar flares are highly energetic, rapidly evolving events in which magnetic energy is converted into thermal, radiative, and kinetic energy \citep{Kowalski24}.
Despite substantial progress in flare modeling \citep{Kerr22,Kerr23}, the broadening of chromospheric spectral lines—especially metal lines such as Mg {\sc ii} h \& k or Ca {\sc ii} H \& K—remains puzzling \citep{Costa17,Zhu19,Kerr24}.
To further complicate the problem, modeling of H {\sc i} Balmer line widths is not nearly as problematic \citep{Kowalski22}.

Multiple processes influence line profiles: Doppler shifts from condensation 
of the chromosphere, Stark broadening from enhanced electron densities, and Zeeman splitting from ambient magnetic fields.
The broadening of hydrogen lines are often relatively well understood, with the primary modeling challenge being matching the Doppler shifts.
But metal lines routinely show much broader Lorentzian wings than expected during strong solar flares that cannot be explained by any line-shape models.

For example, when modeling a solar flare, \citet{Zhu19} found that matching Mg~{\sc ii} wings required artificially multiplying electron impact widths by a factor of $\sim30$.
The resulting profiles were Lorentzian and inconsistent with turbulent broadening unless extremely contrived microturbulence velocity distributions were assumed.
And even when such specific {microturbulent} velocity distributions were assumed in \citet{Zhu19}, it was still a poorer fit than simply increasing the collisional width.
Similar unexplained broadening has been reported in stellar flare observations.

{
Spectral inversions have now been able to reproduce the broad wings in Mg~{\sc ii} using hydrostatic equilibrium atmosphere models, where divergent line of sight velocity gradients, microturbulence, and atmospheric temperature and density structure led to the Lorentzian-like profiles \citep{Sainz23}.
However, it is not clear to what extent the assumption of hydrostatic equilibrium impacts the solution during very dynamic events such as flares. 
As such, we continue to explore alternate solutions. 
}

Following \citet{Zhu19}, we suggest that the mystery of the missing broadening can be resolved {not by atmosphere flows or structure, but by something missing in the intrinsic line broadening.
This would be} either incomplete models of the Stark effect, or that there is another mechanism acting to broaden the line wings.
There have been enough laboratory and theoretical developments that we can rule out the former, so here we explore the latter.
To identify the source of additional broadening, there are two empirical constraints:
\begin{itemize}
    \item The mechanism must broaden metal lines far more than hydrogen lines.
    \item The {intrinsic} line shape\footnote{{ From here on out, ``line shape'' refers to the intrinsic line shape, i.e. the emission/absorption coefficient as a function of wavelength, temperature, and density, which is not affected by radiation transport. This is distinct from the emergent spectra from the sun.}} must be roughly Lorentzian \citep{Zhu19}.
\end{itemize}

Here, we propose that magnetic fields are responsible for the excess broadening.
A single magnetic field produces a clear Zeeman pattern.
Therefore, to create broadening via a magnetic field there needs to be a distribution of magnetic fields in the various layers of the chromospheric plasma to produce broadening that---depending on the distribution---can appear quasi-Lorentzian.

Beyond resolving a long-standing discrepancy in chromospheric metal-line widths, our proposed mechanism has broader implications for flare diagnostics. 
By linking extreme line broadening to localized magnetic structures, it potentially provides a pathway to new avenues for studying phenomena related to flares.
The model further makes a concrete, falsifiable prediction: the presence of a distinctive Stokes V signature in the line wings.

The rest of the paper is as follows.
In section \ref{s:broadreview}, we discuss the recent work on known collisional processes and rule out the possibility that collisional processes can enhance broadening by a factor of $\sim30$ times.
Next, in section \ref{s:Bfields}, we propose that a distribution of magnetic fields can create broad quasi-Lorentzian profiles.
Our model for the probability of magnetic field magnitude, $P(B)$, is empirically created to fit certain criteria.
{In section \ref{s:comparisons}, we present model spectra where magnetically-broadened line-shape profiles are incorporated into RADYN flare simulations (since the broadening is only incorporated into an auxiliary subroutine not part of RADYN); we compare these models against data.}
This proposed explanation is falsifiable, therefore, in section \ref{s:test}, we discuss how flare observations with spectropolarimetry can support or invalidate this idea.
Lastly, we present our conclusions in section \ref{s:conclusions}.
\newline

\section{Processes That Cannot Account For the Broadening}\label{s:broadreview}

None of the many recent advances in flare modeling or spectral line theory can fully explain the broadening observed in flare spectra.
In this section, we examine the different mechanisms we explored as part of the effort to explain the broadening.
{Of the broadening mechanisms discussed, none can supply the apparent factor of $\sim30$ increase in the {Stark-like} broadening {of the wings}.
As discussed in the introduction, it has been demonstrated that spectral inversion techniques result in models that can match data; where relevant in this section we note potential issues with this explanation.
}
\subsection{Microturbulence}

In their conclusions \citet{Sainz23} state that in their inversion fits, ``the extended, broad wings of the Mg~{\sc ii} h \&k are sensitive to changes in $\nu_{\rm{turb}}$ in the mid chromosphere''.
Microturbulence usually manifests as Doppler broadening, {yet the emergent spectral} profiles are roughly Lorentzian in shape.
Therefore, if microturbulence is the explanation, then a specific $\nu_{\rm{turb}}$ structure is required to create the Lorentzian shape.

We empirically note that this roughly Lorentzian shape is fairly consistent across different flares{, especially stronger flares}.
Using the unsupervised machine learning algorithm \textit{k}-means to classify spectral line profiles, \citet{Panos18} show commonalities between different types of flares, one of which is the consistency of broad Lorentzian profiles in { stronger flares, such as those exemplified in \citet{Zhu19}}.
{Looking for alternate explanations,} \citet{Zhu19} simulated different $\nu_{\rm{turb}}$ structures required to reproduce the observations {using RADYN simulations}, increasing with depth into the chromosphere {(without enhancing the broadening)}.
There, the magnitude of $\nu_{\rm{turb}}$ is much larger than the predictions of \citet{Sainz23} and is inconsistent with other observations such as O~{\sc i} \citep{Kerr24}. 
Therefore, we don't expect that the variation of different $\nu_{\rm{turb}}$ structures based on different initial conditions of the flare would result in consistent Mg~{\sc ii} line shapes across different flares.

\subsection{Underestimation of the Chromospheric Density}\label{s:overdensity}

One possible explanation for the missing broadening is that the average density in the compressed chromosphere is larger than expected by RADYN simulations, thus increasing the broadening.
{In their conclusions \citet{Sainz23} further discuss the temperature and density structure in the chromosphere, stating that the density ``reaches its maximum in the low chromosphere with a value of $\log (n_e)\approx 15$, just before the temperature minimum, which is pushed down (in terms of the optical depth) towards the low-photosphere $(\log\tau\approx -1)$.''.
However, at this layer in the atmosphere, the inversion results are unreliable.
A density of $\log (n_e)\approx 15$ is not much higher than the chromospheric densities predicted by RADYN. 
However, the large densities predicted by RADYN are higher up in the atmosphere than $\log(\tau)\sim -1$.
When \citet{Sainz23} perform their inversion with other lines, the chromospheric and photospheric densities are reduced substantially, being consistently under $\log(n_e)\gtrsim 13$ throughout the entire atmosphere, reaching $\log(n_e)\approx 14$ in some instances, but only in the lower parts of the atmosphere, not in the chromosphere.
It is not immediately clear why the RADYN-like stratification differs from the inversions, but a detailed comparison should be performed.
}

The idea of larger densities should be considered more fully; could the average density in the compressed chromosphere be underestimated by a factor of $\sim$30?
We find this to be highly implausible {as the enhancement in the electron density would be at least 3,000 times that of the quiet sun chromosphere as this is inconsistent with both RADYN simulations and atmosphere structure from inversions.}
Such a large density enhancement would be immediately obvious in the hydrogen Balmer lines, which are extremely sensitive to density.
Such correspondingly broad Balmer-series lines are not generally detected in spectroscopic measurements \citep{Costa16,Tamburri26}.

There may be some stratification in the atmosphere, and lines can form at different layers.
In this picture hydrogen could form at less dense layers, while Mg could be coming from a 30x compressed layer.
While this might be the case, the complete separation of the hydrogen and Mg where the latter has a density 30 times that of the latter seems unlikely.

{\subsection{Plasma Flows}}

The magnitude of flows predicted {by the \citet{Sainz23}} inversions is not uncommon in radiation hydrodynamic simulations \citep[e.g.][]{Kowalski17}, but these simulations do not predict the pointy profiles. 
It is not clear, then, why line profiles with extremely broad Lorentzian wings are not routinely produced by flare models \citep[e.g.,][]{Zhu19}. 
Important to note is that modeling a very dynamic system under the approximation of hydrostatic equilibrium can be problematic, which could impact the density (therefore optical depth) structure of the flare atmospheres. 
As noted by \citet{Kerr24}, direct comparison of forward modeled dynamic flare atmospheres with inversions, with an appreciation of the assumptions in each approach, is critical in understanding this discrepancy.

\subsection{Pressure Broadening}
In a dense plasma, the radiating atom experiences collisions from the plasma electrons as well as stray microfields from nearby ions.
Various models are able to account for different physical effects.
Some considerations include penetrating collisions by both ions and electrons, wherein penetrating collisions are where plasma particles enter into the radial extent of the radiating atom's wavefunction \citep{Woltz84,Junkel2000,Gomez21,Stambulchik22}.
Accounting for the penetration generally results in corrections to the classic Stark broadening models \citep{Gomez24PRA} resulting in a reduction of the width.

Because ions move much more slowly than electrons, then they are often treated as static. 
It has been shown that accounting for ion dynamics can result in significant corrections for some lines \citep{Gigosos87,Stambulchik06,Gomez16}, most notable in $\alpha$ transitions of hydrogenic lines, such as Ly$\alpha$.
Ion broadening, whether static or dynamic, does not affect lines where levels are isolated \citep{Alexiou14}, such as Mg~{\sc ii} and Ca~{\sc ii}, as we confirmed using the {\sc Xenomorph} code \citep{Cho22}.

Even though the physics of collisions is well understood, first-principle calculations are discrepant with experimental line width measurements \citep{Glenzer92,Glenzer96,Blagojevic99}.
The reason for this discrepancy---the magnitude of the which never exceeds a factor of two \citep{Griem97,Ralchenko03} not 30 as suggested by \citet{Zhu19}---remains a mystery.

\subsection{Collisions From the Electron Beam}
High energy particles from the outer atmosphere are accelerated toward the solar surface. 
These non-thermal high energy particles form a beam that deposit energy to the chromosphere.
\citet{Hawley07} suggested that the beam could enhance the collisional rates.
We examined this from two perspectives: direct collisions from the beam, and secondary effects, such as plasmon excitation.

We found that non-thermal electron beams, with densities $\sim10^{8}-10^{10}~\mathrm{cm}^{-3}$ and energies $\gtrsim20$ keV \citep{White03, Krucker11,Volpara25}, cannot supply the excess broadening through direct collisions.
The reason for this is two-fold.
First, slow electrons usually provide the largest perturbation.
Faster electrons don't have time to interact with the radiating electron to produce a significant perturbation.
There might be some small effect regarding inner shell ionization, but these processes are much weaker compared to collisional excitation between levels of the same principal quantum number.

Secondly, the density of the beam is far too low to compete with thermal electron collisions in the chromosphere.
The density of the compressed chromosphere is estimated to be of order $10^{14} ~\mathrm{cm}^{-3}$, at least 4 orders of magnitude larger than the beam density.
The beam just cannot compete.

It is well established that high energy particles that 
impact into a material can excite plasmon (plasma wave or Langmuir wave) modes \citep{PinesIV}.
Beam-generated Langmuir waves \citep{BohmGrossA,BohmGrossB,BohmPinesIII,PinesIV} create detectable radio signatures and hydrogenic satellites \citep{BarangerMozer61,Hannachi23}.
These satellites in the wings of spectral lines usually appear at the plasma frequency away from line center.
We have previously found that plasmon waves have little effect on the isolated metal lines of interest like Mg~{\sc ii} h\&k \citep{GomezPlasmon25}.

Plasmon waves can also modify free electron trajectories during collisions.
While the theory for this (to our knowledge) has not been developed, it is not expected that this can create factors of 30 enhancements.


\begin{figure*}[t!]
\centering
\includegraphics[scale=0.70,trim=0cm 0 0 0]{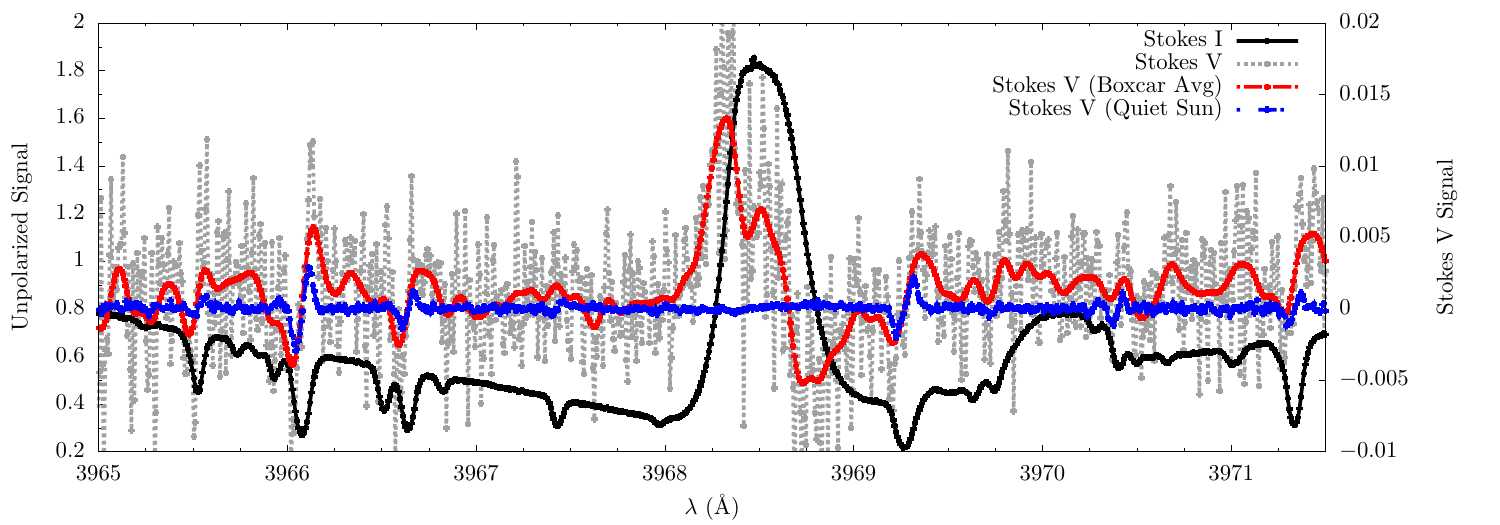}
\caption{Spectropolarmetric data of Ca~{\sc ii} H 3968 \AA\ during the decay phase of the C6.7 class flare on 19 August 2022, 20:42UT using DKIST/ViSP.  The spectral window also includes the Balmer-series H$\epsilon$ line at 397.0 nm.  For details of the observing configuration, see \cite{Tamburri26}. Black corresponds to unpolarized signal, grey is the Stokes V signal, red is a boxcar average, and the blue  line is a quiet-sun spectrum for reference. The increased Stokes V signature in the ribbon indicates an increase in magnetic fields.
}
\label{DKISTdata}
\end{figure*}

{
\section{Broadening from  a Magnetic Field Distribution}\label{s:Bfields}
}

Flares necessarily occur in regions of strong, complex magnetic fields from photospheric to coronal heights \citep{FlareBook}. 
In spectropolarimetric observations of a C-class flare from the Visible Spectropolarimeter at the Daniel K. Inouye Solar Telescope \citep[DKIST/VISP ][]{DKIST,VISP}, \citet{Yadav25} showed 500G fields that appeared shortly after the impulsive phase.
In Fig. \ref{DKISTdata}, we show spectropolarmetric data from DKIST/ViSP  during the decay phase of a  C6.7 class flare on 19 August 2022 \citep{Tamburri26}.
The Ca II H Stokes V signal is much stronger in the flare ribbon (red line) than in the quiet-Sun spectrum (blue line).
This increase in the Stokes V signal indicates that there is increased magnetic activity compared to the quiet sun.

We propose that magnetic fields present in the flare atmosphere may be responsible for the anomalous broadening of metal lines.

Throughout this work, electromagnetic quantities are expressed in SI units unless explicitly stated otherwise.
Magnetic field strengths are occasionally quoted in Gaussian units for observational context, with
$1~\mathrm{T} = 10^4~\mathrm{G}$.
Particle densities are number densities.
Astrophysical quantities, such as electron beam flux, are given using standard cgs units.
In appendix \ref{BalrogUpgrade}, where we discuss details of an atomic spectral model, atomic units are used.

\subsection{ Empirical Magnetic Field Distribution}

For magnetic fields to explain the broadening, a distribution $P(B)$, of field strengths is required.  The final spectrum is obtained by integrating over this distribution.
If magnetic fields are to explain the extreme quasi-Lorentzian broadening of Mg~{\sc ii} lines, $P(B)$ must {produce a line shape that's roughly Lorentzian.}
For this to be the case, the line-shape function, i.e. the intensity as a function of { angular} photon frequency, should scale as $I(\omega)\sim \omega^{-\alpha}$, where $\alpha$ is between 2 and 3.
For the metal lines of interest, the $\omega^{-\alpha}$ requirement directly translates to a $P(B)$ that has a $B^{-\alpha}$ power-law tail; why this is the case will be elaborated below.

\begin{figure}[h!]
\centering
\includegraphics[scale=0.70,trim=0cm 0 0 0]{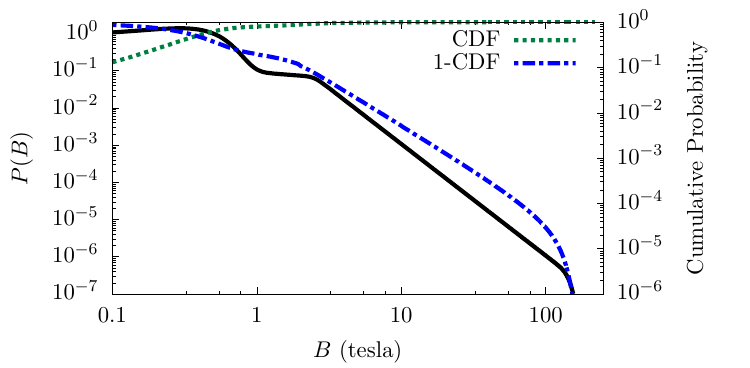}
\caption{An example $P(B)$ distribution for an extremely strong flare with $B_{max}=150$T with $\alpha=3$. In this model, the probability of large magnetic fields occurring compared to the background in the flare ribbon is less than roughly 1 part per million. 
Shown also is the cumulative distribution function (CDF, dotted green), where it is seen that the CDF goes to unity very quickly as the field increases beyond the background.
Also shown is 1-CDF (dot-dashed blue), to show the shrinking probability of higher fields.
}
\label{PB}
\end{figure}

To reflect a more physical system, our empirically-driven model for $P(B)$ contains a background magnetic field dominated by the field in the ribbon plus this power-law behavior with a smooth transition between the filament field with small $B$ and large $B$ roll-offs,
\begin{multline}
    P(B) \propto \exp(-\frac{1}{2}(B-B_{\mathrm{bg}})^2/\sigma_{B_{\mathrm{bg}}})\\
    + f\left[\frac{B}{B_{\min}}\right]^{-\beta}\Bigg[1+\left(\frac{B}{B_{\min}}\right)^{1/\delta}\Bigg]^{(\beta-\alpha)\delta}\times\\
    \frac{1}{2}\mathrm{erfc}\left(c\frac{B-B_{\max}}{B_{\max}}\right),
    \label{e:PB}
\end{multline}
where $B_{\mathrm{bg}}$ is the background field with a corresponding Gaussian width $\sigma_{B_{\mathrm{bg}}}$.
The background field, weighted by $f$ (which is of order 0.01-0.03), has a high-energy tail that extends to $B_{\rm max}$ which is rolled off using the complementary error function. 
{$B_{\rm min}$ will generally be of order $B_{\rm bg}$, maybe a factor of two larger.}
$c$ is a dimensionless parameter that controls the slope of the high-field cutoff.
$P(B)$ is divergent if $B^{-\alpha}$ is extended to arbitrarily small fields. We therefore choose a functional form that is designed to roll off $P(B)$ at small $B$.
$\beta$ and $\delta$ are chosen to dictate the small $B$ behavior.  For the models presented here, $\beta=0.3$ and $\delta=0.05$. 
An example of this field distribution is shown in Fig. \ref{PB}, where we also show the cumulative distribution.

The exact choice of $B_{\rm bg}$ is unimportant for Mg~{\sc ii} h\&k. 
This is because even at fields above 0.3T, the splitting is not enough to overcome the contribution from thermal Doppler broadening.
Therefore, for the analysis below, we will not present any definitive values for $B_{\rm bg}$.

\begin{figure*}[ht!]
\centering
\includegraphics[scale=0.73,trim=0cm 0 0 0]{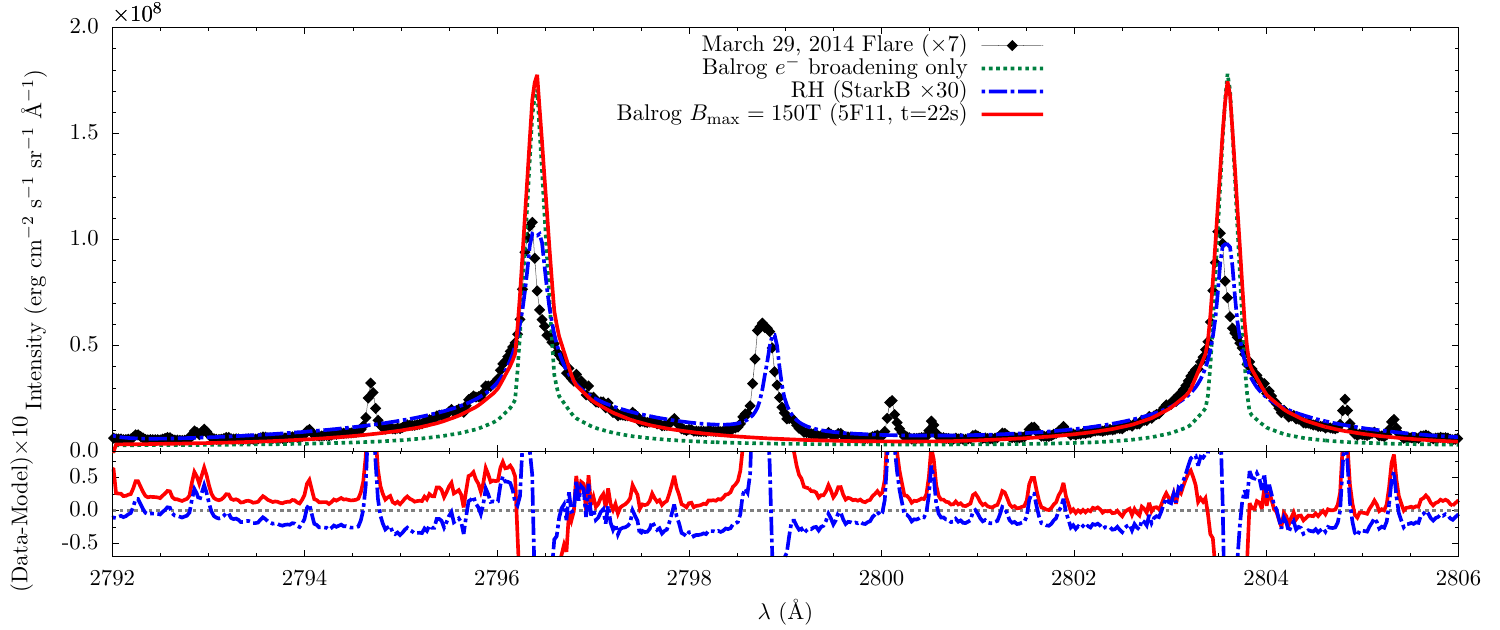}
\caption{Model comparisons of the X1.1 flare from March 29, 2014 (not $\chi^2$ minimized). The observations have been scaled by a factor of 7 to match the Balrog models in the continuum in the far wing of Mg~{\sc ii} h. Dotted green is without the magnetic field, solid red includes magnetic field broadening from the $P(B)$ in Fig. \ref{PB} using $\alpha=3$, and dot-dashed blue is the x30 model used in \citet{Zhu19}.
The magnetic field clearly not only adds broadening, but also fits the wings of the lines far closer than the x30 Lorentzian model.
We ignore the goodness of fit in the core of the line (elaborated in the text).
The difference between data and model is shown in the bottom panel. It's clear that even though the far-wings are comparable, the Lorentzian ($\omega^{-2}$ behavior) over-predicts the wings compared to the $\alpha=3$ magnetic-broadening model ($\omega^{-3}$). Further, in the far wings, the data-model plot of the $\alpha=3$ model is flatter than the Lorentzian.
}
\label{2014_3_29}
\end{figure*}

Since the full width at half maximum of the electron collisional width is much smaller than the thermal Doppler broadening, the line profile at line center will be roughly Gaussian for each magnetic field value
\begin{equation}
    I_0(\omega,B)\propto \sum_i\exp(-\frac{1}{2}(\omega-\omega_{0,i}(B))^2/\sigma^2)\,;
\end{equation}
here $\omega$ is the {angular} frequency of radiation {(where the energy of the photon is $E=\hbar \omega$), and the index $i$ corresponds to each Zeeman component of the spectral line}.
Since the splitting of the field is roughly linear and slowly varying (see details for how shifts in energy from magnetic fields are calculated in Appendix \ref{BalrogUpgrade}), then the final line shape---obtained by an integral over $P(B)$,
\begin{equation}
    I(\omega) = \int_0^\infty dB P(B)I_0(\omega, B)
\end{equation}
---will scale as how $P(B)$ scales with magnetic field.
In other words, of $P(B)\sim B^{-3}$, then the line wings will likewise scale as $\omega^{-3}$ {because $\omega_{0,i}$ is roughly linear in $B$}.
Therefore, $\alpha=2$ distributions will be indistinguishable from a true Lorentzian, but an $\alpha=3$ distribution will be distinct from a Lorentzian and the profile will be slightly steeper.

\subsection{Consequences of this Magnetic Field Distribution}

A point of interest with this model is how such a distribution affects the magnetic energy budget.
For this idea to be plausible, the high-energy tail cannot significantly raise the magnetic energy density over the background field.
The magnetic energy density of the background field is given by 
\begin{equation}
    u_B = \frac{B_{\rm bg} ^2}{2\mu_0}.
\end{equation}
Inclusion of the high-$B$ tail means that the average magnetic energy density acquires an additional term when integrating over $P(B)$.
The addition to the magnetic energy density is the log of the ratio of the $B_{max}$ to the background,
\begin{equation}
    \Av{u_B} \approx \frac{B_{\rm bg}^2}{2\mu_0} + f\frac{B_{\rm bg}^2}{\mu_0}\ln\frac{B_{\rm max}}{B_{\rm bg}},
    \label{UB}
\end{equation}
The amount of additional energy from the power-law tail adds only 10-15\% to the total energy budget compared to the background field even when $B_{\rm max}= 100$T.
{

Further, this does not significantly alter the average $B$ at a given layer in the atmosphere.
Assuming that $B_{\rm max}\gg B_{\rm min}$, we derive that
\begin{equation}
    \langle B\rangle =\int_0^\infty dB~P(B) B \approx (1-f)B_{\rm bg} + fB_{\rm min}.
\end{equation}
If $B_{\rm min}= 0.5T$ and $f=0.03$, then the average $B$ only increases by 150G, regardless of $B_{\rm bg}$.
}


{
\subsection{Possible origin of the field distribution}
The magnetic energy density constraint suggests that the origin of such fields is unlikely to be directly associated with the electron beam itself. 
While return-current physics \citep{Knigh77,vandenoord90} and beam filamentation \citep{Bret05,Tzoufras06,Fiuza20,Raj20,FiuzaPrivate} may produce localized current structures, these are typically limited to field strengths of order $\sim 10^2\,\mathrm{G}$ under flare conditions and are therefore insufficient to account for the extreme tail.

We therefore speculate that the relevant magnetic fields are better understood as being amplified by plasma compression under flux-freezing conditions. 
In this picture, magnetic flux is approximately conserved throughout the chromospheric dynamics, so that local increases in density during flare-driven compression naturally lead to corresponding enhancements in magnetic field strength.
For the fields to respond to compression, they would have to be anti-aligned with the vertical plasma flows during compression.

Taking a representative pre-flare initial chromospheric field strength of $B_{\rm init} \sim 100\,\mathrm{G}$ and electron density $\langle n_e\rangle_{Av} \lesssim 10^{12}\,\mathrm{cm^{-3}}$, X-class flare conditions can drive localized compressions to $\langle n_e\rangle_{Av} \sim 1$--$5\times 10^{14}\,\mathrm{cm^{-3}}$ \citep{Kowalski26}, corresponding to field strengths of order $1$--$2\,\mathrm{T}$ under this heuristic scaling. 
The extreme tail of the distribution would require transient overdensities significantly above typical flare condensations, occurring in regions with filling factors $\lesssim 10^{-6}$.

Such structures are expected to be highly transient rather than equilibrium configurations, forming in response to impulsive and spatially inhomogeneous energy deposition and rapidly relaxing through expansion, shocks, and radiative losses. 
The resulting high-field tail should therefore be interpreted as a statistical manifestation of a strongly time-dependent and spatially intermittent flare atmosphere, rather than evidence for a persistent magnetic configuration.

The details of the flare dynamics that generate this distribution are not addressed here. 
We emphasize that this work does not attempt to derive the magnetic field distribution from first-principles flare MHD, but instead adopts a phenomenological model designed to be consistent with basic energetic constraints and to yield testable predictions for the enigmatic line broadening of Mg~{\sc ii} lines.

}


\section{Comparison of Model with Flare Spectra}\label{s:comparisons}

Using flare atmospheres computed with RADYN, we calculated line profiles with our upgraded version of {\sc Balrog} \citep{Gomez21,GomezLyA}, a semi-analytic spectral line shape code that treats electron collisions quantum mechanically.
{\sc Balrog} now includes fine structure and multi-electron effects (Appendix~\ref{BalrogUpgrade}).
RADYN \citep{Carlsson97,Allred05,Allred15} is a 1D plane parallel radiation hydrodynamics (RHD) code that calculates heating from a non-thermal electron source.
It solves the non-local thermodynamic equilibrium (NLTE) radiative transfer equations and the hydrodynamic equations assuming the electron beam propagates downward along a 1D flux tube.  
The Mg {\sc ii} spectra are calculated assuming the Mg population densities in local thermodynamic equilibrium (LTE).  
{ The subroutine to generate the spectra follows the method presented in \citet{Kowalski17}.}
The other atmospheric variables (gas density, electron density, temperature, gas velocity, and hydrogen population densities) are obtained from specific snapshots in three RADYN models from \citet{Kowalski22} and \citet{Kerr24}.  
Given the Mg {\sc ii} line and continuum opacities and emissivities at each height, the emergent intensities are calculated using a Feautrier solver from the RADYN source code.

In this model, $B_{\mathrm{bg}}$ and the corresponding tail are assumed to be constant across the entire line-forming region of Mg~{\sc ii} h\&k.
Ideally, forward modeling would have $B$-field structure as a function of height in the atmosphere.
We point out that the high-$B$ tail in our model would never be produced by inversion techniques \citep[e.g. ][]{AllendePrieto01} using a 1D atmosphere.
Therefore, such a distribution would only be possible through 3D modeling of the solar atmosphere such as in \citet{Asplund2000} or \citet{Rempel23}, where at every height in the atmosphere there is a range of temperatures, densities, and magnetic fields.

Our first comparison is with the X1.1 flare from March 29, 2014, previously studied by \citet{Zhu19} and observed by IRIS \citep{IRIS}.
To reiterate, the \citet{Zhu19} study used the electron broadening $\times$30 model to match the data.
The {\sc Balrog} results (natural, electron broadening, and thermal only) without the magnetic field broadening, like \citet{Zhu19}, demonstrate that it is woefully unable to match the line wings.
With the $P(B)$ used in Fig. \ref{PB}, the {\sc Balrog} calculations are able to more closely match the wings of the spectral lines.
This model matches better than the x30 model of \citet{Zhu19} in the far line wings.

Some mismatch remains in the core, likely reflecting uncertainties in the ratio of background fields to filament fields or limitations of the 1D atmospheric model.
Additionally, the assumption of LTE can impact the intensity in the core.

\begin{figure}[t!]
\centering
\includegraphics[scale=0.26,trim=0cm 0 0 0]{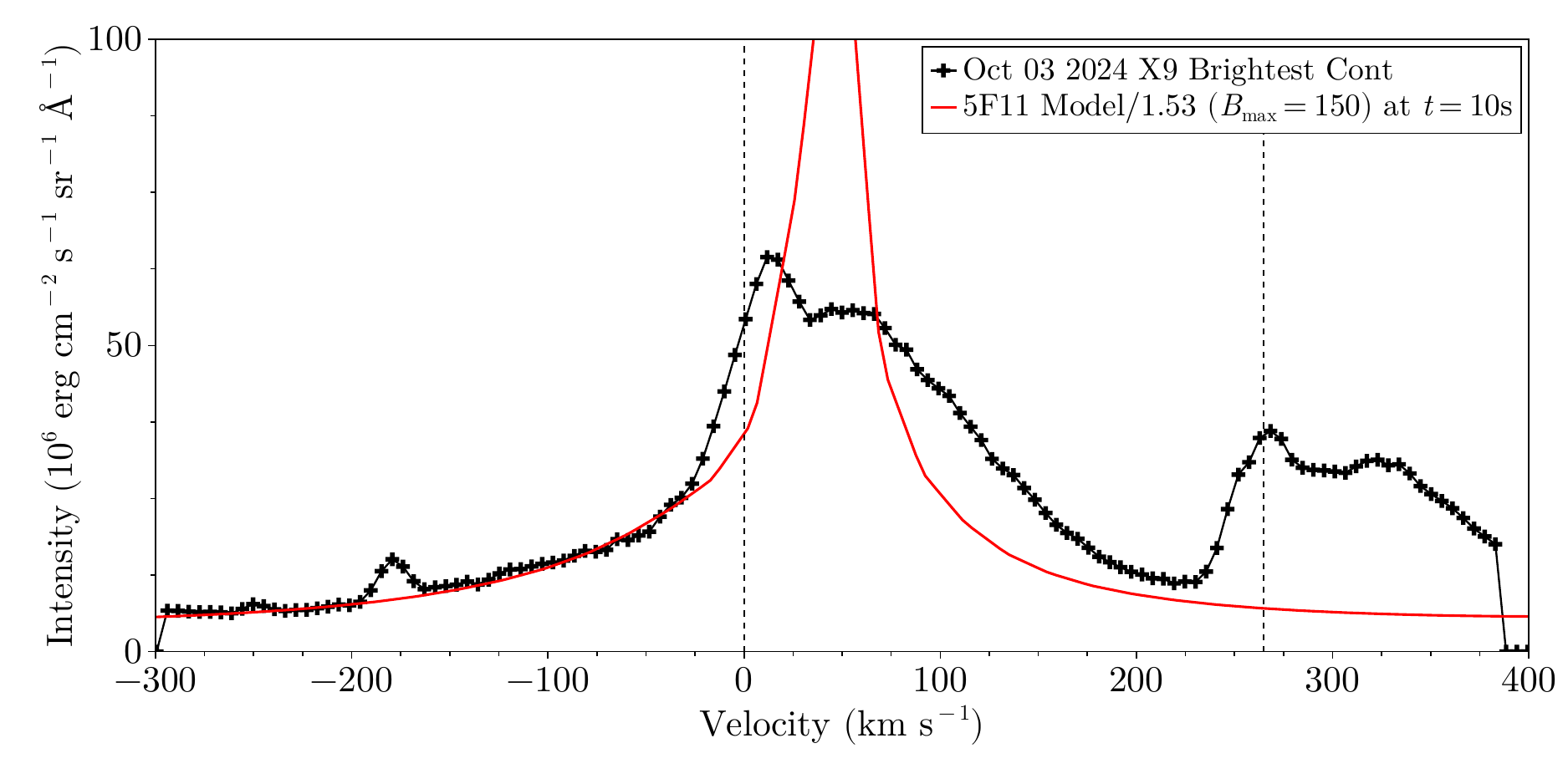}
\caption{Comparison of the new broadening model ($\alpha=3$) against the X9 flare on Oct 3, 2024. This flare has much stronger Doppler shifts than what is present in Fig. \ref{2014_3_29}, yet, while the Doppler shifts significantly change the core of the line, the blue wing especially is  well matched by the magnetic broadening.
}
\label{2024}
\end{figure}

In Fig. \ref{2024}, we examine another flare, specifically an X9 flare from Oct 3, 2024.
During this flare, the chromosphere exhibits much more severe Doppler shifts not captured by the RADYN model.
Yet, despite this, the blue wing profile (which is less susceptible to Doppler broadening, since the mass flow is directed towards the chromosphere, away from the observer) matches extremely well with the magnetic field broadening model.

\begin{figure}[!]
\centering
\includegraphics[scale=0.26,trim=0cm 0 0 0]{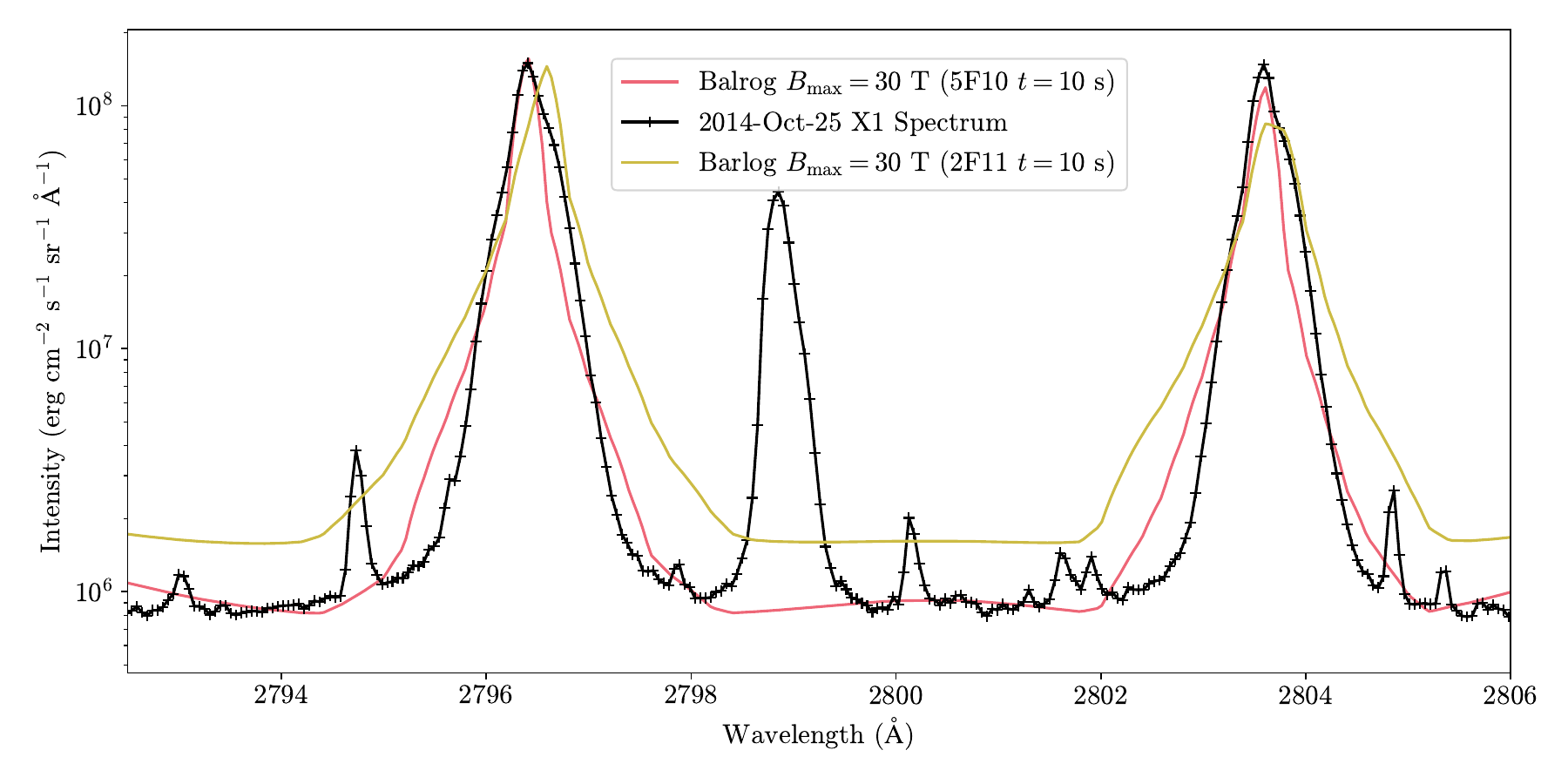}
\caption{This is a model comparison ($\alpha=3$) against the X1 flare on Oct 25, 2014. This spectrum does not exhibit the same broadening as either the March 29, 2014 or Oct 3, 2024 flares. 
This is where the $B_{\max}$ parameter of the model plays a role as the profiles begin broad, but are dampened part way in the wings, which is readily apparent in the log intensity.  The observed spectrum is in DN/s/pixel, and an arbitrary uniform scale factor has been applied.
}
\label{2014_10}
\end{figure}

We also compare this model against the X1 flare on Oct 25, 2014, previously analyzed by \citet{Kerr24}.
This event was complex, with multiple episodes of non-thermal electron injection (inferred from multiple bursts of hard X-rays). The sources studied by \citet{Kerr24} occurred during the last, and weakest, of these HXR bursts though the analysis of hard X-rays \citep[see also][]{Kowalski19} still suggested the potential of a large energy flux density as an upper limit, ranging $F=2\times10^{10} - 2\times10^{11}$~erg~s$^{-1}$~cm$^{-2}$.
The broadening is not quite as intense as in the other flares, and indeed, examining the spectra---shown in Fig. \ref{2014_10}---the wings are not nearly as extended as in Figs. \ref{2014_3_29} or \ref{2024}. It is not known why this is the case, though it is interesting to note that as well as the fairly small lower limit of $F$, a rather soft non-thermal electron distribution was obtained from the X-ray analysis, with $\delta \sim8$.
The model that includes a $B_{\rm{max}}$ cutoff fits this model quite well as the broadening near the core is large, but then decays rather quickly beyond a certain detuning. 
The smaller broadening, and subsequent difference to the {\sc Balrog} modeling demonstrates both the versatility of our proposed model, as well as its potential as a diagnostic.

\section{An Observational Test}\label{s:test}

A key observational prediction for this model is the polarization signature. 
If indeed the additional broadening is a result of $B$ fields the signature would be unambiguously evident in the Stokes V profile, as seen in Fig. \ref{DKISTdata}: the wings on one side would be positive and negative on the other.
If the anomalous broadening were due to collisional process or some (microturbulent) velocity, then the signal would be weaker and the wavelength extent of positive and negative areas of the Stokes V measurements would be much smaller, if not non-existent, and would not extend into the wing.
Therefore, the lack of a polarization signature in the Stokes V would favor alternate explanations over Zeeman broadening, such as microturbulent velocities and strong velocity gradients \citep{Sainz23}.
However, if Zeeman broadening is indeed the source, then spectropolarmetric observations of these lines would reveal a strong {\em and wide} Stokes V signal, probably comparable to the unpolarized signature.
{
In other words, if magnetic broadening dominates, the fractional polarization in the line wings, $|{\rm Stokes}~ V/{\rm Stokes}~ I|$, would be of order unity and symmetric about the line center; this contrasts with the weak-field limit, where $V \propto B\,\frac{dI}{d\lambda}$ and $|V/I| \ll 1$.
This latter condition implies little to no magnetic broadening.
}
Detecting this predicted Stokes V signal in extremely broadened metal lines during  solar flares could confirm the validity of this model.

Zeeman broadening only shows up in the metal lines such as Mg~{\sc ii} h\&k and Ca~{\sc ii} H\&K because the pressure broadening is so weak.
The pressure broadening for H is far stronger, and therefore, the Zeeman broadening would only broaden H lines minimally, if at all.

However, there is a caveat with this broadening signature appearing in metal lines. 
We emphasize that since the wings are broadened by a low-probability high-$B$ tail, then this effect would be the most distinct in saturated spectral lines.
Weak metal lines would not exhibit this additional broadening.
As was demonstrated in \citet{GomezLyA}, when spectral lines are saturated (as is the case with Ly$\alpha$ in white dwarfs), then changes in the profile that are one millionth the intensity of the peak of the line appear in the emergent spectrum.
The explanation for $B$ fields affecting the wings has this exact same behavior, it occurs so far down the profile in terms of intensity, but only appears in the spectrum because these lines are saturated.


Therefore, the best chance of observing this signature would be spectropolarimetry of Mg~{\sc ii} h\&k, which is at the intersection of being susceptible to magnetic field broadening and is saturated enough to show it.
But the only instrument capable of observing h\&k is IRIS, and IRIS does not have a spectropolarimeter.
The next best candidate would be the Ca~{\sc ii} H\&K lines, which are observable in full-Stokes spectropolarimetric observations from DKIST/ViSP.

\section{Conclusion}\label{s:conclusions}

During solar flares, the broadening of some spectral lines are much larger than can be predicted by currently-known broadening mechanisms.
We propose that this anomalous and enigmatic broadening is due to Zeeman broadening from a distribution of magnetic fields, where the wings are broadened by high-$B$ power law of the distribution, i.e. $P(B)\sim B^{-\alpha}$.
An $\alpha=3$ matches observational datasets best.

Verification of this model is possible with spectropolarmetric observations of solar flares.
The Zeeman broadening produces a {\em strong} asymmetric feature in the Stokes V filter. Neither Doppler shifts or pressure broadening can produce this signature.
Therefore, observation of a strong distinct asymmetric Stokes V signal would verify magnetic fields being responsible for the broadening. 
The DKIST presents an ideal observatory to hunt for this signal, and we strongly advocate for dedicated spectropolarimetric flare campaigns (e.g. focusing on Ca{~\sc{ii}} H \& K). 
Additionally, the Chromospheric Magnetism Explorer \citep[CMEX; ][]{cmex} is a proposed mission concept that would provide spectropolarimetric observations of the Mg~{\sc{ii}} h \& k lines.


\section{Acknowledgments}

We thank Dr. J. Allred and Prof. F. Fiuza for valuable discussions.
T.A.G. acknowledges support from the George Ellery Hale Post-Doctoral Fellowship at the University of Colorado, the NNSA Stockpile Stewardship Academic Alliances under Grant No. DE-NA0004100, and the U.S. Department of Energy NNSA Center of Excellence under cooperative agreement number DE-NA0004146.
GSK acknowledges NASA's Early Career Investigator Program (grant \# 80NSSC21K0460) and from the Royal Society via a University Research Fellowship (grant \# URF/R1/251160). JRW acknowledges support from the United States Department of Energy National Nuclear Security Administration SSGF program under DE-NA0003960, and the U.S. Department of Energy NNSA Center of Excellence under cooperative agreement number DE-NA0004149. LANL is operated by Triad National Security, LLC, for the National Nuclear Security Administration of the U.S. Department of Energy under Contract No. 89233218NCA000001.
IRIS is a NASA small explorer mission developed and operated by LMSAL with mission operations executed at NASA Ames Research Center and major contributions to downlink communicaitons funded by ESA and the Norwegian Space Centre.

\appendix

\section{Upgrades to the Balrog Code for Multi-Electron Radiators}\label{BalrogUpgrade}

The Balrog code \citep{Gomez21,GomezLyA} has already been developed for the accurate broadening of one-electron atoms.
For this work, we are not interested in the far line wing ($\Delta\omega\gg FWHM$) behavior of the spectral lines, therefore we will neglect the impact of the transient terms of electron broadening.

The classic expression for the spectrum involves an integral over electric microfields,
\begin{equation}
    I(\omega) = \int_0^\infty d\varepsilon W(\varepsilon)J(\omega,\varepsilon)
\end{equation}
where the distribution function $W(\varepsilon)$ is easily calculated by codes such as APEX \citep{Iglesias84}, and $J(\omega, \varepsilon)$ is the electron-broadening function
\begin{equation}
    J(\omega,\varepsilon) = \frac{-1}{\pi}\mathrm{Tr}D\frac{1}{\omega-L_a(\varepsilon) - \EB(\omega)}\rho_aD,
\end{equation}
where $D$ is the atomic transition dipole moment, $L_a(\varepsilon)$ is the atomic Liouville operator that's dependent on the field values, defined as 
\begin{equation}
    L_a(\varepsilon)\rho = [H_a(\varepsilon),\rho] = H_a(\varepsilon),\rho - \rho H_a(\varepsilon),
\end{equation}
and $\EB(\omega)$ is the electron-broadening operator and $\rho_a$ is the density matrix of the atomic states, usually taken to be a Boltzmann distribution.
These operators have four indices, rather than the usual 2, and $D$ is functionally a vector.
The electron-broadening operator is given by 
\begin{equation}
    \mat{ab}{\EB(\omega)}{a'b'} 
\end{equation}
where $a$ and $b$ are a set of upper and lower states, while $\mat{a}{D}{b}$ connects those upper and lower states.

The electron broadening is given as a sum of collision $T$-matrices:
\begin{widetext}
    \begin{multline}
\mat{ab}{\EB(\omega)}{a'b'} \approx n_e\lambda_T^3 \int_0 dk k^2 e^{-\beta E_k}\times
    \Bigg\{\mat{ak}{T(E_b+E_k+\omega)}{a'k}\delta_{bb'}
    - \delta_{aa'}\mat{bk}{T^*(E_a+E_k-\omega)}{b'k}\\
    + i\pi \int dk' k'^2 \times
    \bigg[\delta(E_{a}+E_{k}-\omega-E_{b'}-E_{k'})\mat{ak}{T(E_{a}+E_{k})}{a'k'}\mat{bk}{T^*(E_{b'}+E_{k'})}{b'k'}\\
    +\delta(\omega-E_{a'}-E_{k'}+E_{b}+E_{k})\mat{ak}{T(E_{a'}+E_{k'})}{a'k'}\mat{bk}{T^*(E_{b}+E_{k})}{b'k'}\bigg]\Bigg\},
    \label{eq:ebroad}
\end{multline}
\end{widetext}
where the first term is the broadening of the upper state, the second term is the broadening of the lower state, and the last term is the interference term.
Physically these last two terms are corrections to the broadening that account for the fact that the collision process is not independent processes.
In narrow metal lines, especially those of $\Delta n=0$ such as what are considered here, the corrections due to correlated collisions are substantial \citep[e.g.][]{Griem97}.

For this work, the elements of interest are what are known as isolated-line transitions.
Isolated lines are lines that (except for fine structure splitting), have energy levels that are far apart.
These broadening of these transitions are largely driven by electron impact collisions and ions tend to play a fairly minor role; \citet{Alexiou14} states ``for isolated lines, ions normally cannot compete with electrons" and lists the exception as being for quadrupole interactions, but this is still a relatively small correction.
We verified this using the simulation code \X~\citep{Gomez16,Cho22}.
Therefore, we will neglect the effect of the ions.

Since the problem with flares involves magnetic fields, we will instead integrate over magnetic field distributions.
This modifies our expression so that the spectrum is 
\begin{equation}
    I(\omega) = \int_{B_{\rm min}}^{B_{\rm max}} d\beta P(\beta)J(\omega,\beta)
\end{equation}
where
\begin{equation}
    J(\omega,\beta) = \frac{-1}{\pi}\mathrm{Tr}D\frac{1}{\omega-L_a(\beta) - \EB(\omega)}\rho_aD.
\end{equation}
To evaluate the impact of the magnetic field on the spectrum, we include only the linear Zeeman effect, which is the operator
\begin{equation}
    H_{mag}(\beta) = -\frac{1}{2}\beta\mu_0(\mathbf{J}+(g_s-1)\mathbf{S}),
\end{equation}
where $g_s\cong$2.0023192.
Here, $\beta$ is the magnetic field, $\mu_0 = 4.2543\times10^{-10}$Ry/gauss, and the matrix elements of the atomic operators is diagonal in all quantum numbers except $J$,
\begin{widetext}
\begin{equation}
 \frac{2}{\beta\mu_0}\mat{a}{H_{mag}}{a'} = M\delta_{aa'} - \delta_{\alpha L S M,\alpha'L'S'M'}(g_s-1)(-1)^{L+S+M}[J,J']^{1/2}\{S(S+1)(2S+1)\}^{1/2}\threeJ{J}{1}{J'}{-M}{0}{M}\sixJ{L}{S}{J}{1}{J'}{S}
\end{equation}
\end{widetext}
where we have assumed the field is aligned in the $z$-direction.
The linear Zeeman operator is diagonal in all quantum numbers except $J$.

For this work, we will ignore the impact of the magnetic field on the collision problem and will only account for it in the final step of the calculation.
For the average magnetic fields in this problem, accounting for the spiraling trajectories as was done in \citep{Gomez23,Gomez24} is both unnecessary and an onerous task.
Only in the highest magnetic fields, such as those exceeding a MG, would accounting for the quantization of the perturbing electron into Landau levels be necessary.
Our aim is to understand the effect of the magnetic field from the flare beam, and if further refinement of the electron broadening is needed, then that will be reserved for future work.

\subsection{Evaluation of $T$-matrices}
For this work, we solve the $T$-matrices using linear solve (i.e. $Ax=b$) routines such as those found in LAPACK.
The solution for the $T$-matrix is 
\begin{equation}
    [1-V(E-H_0)^{-1}]T(E) = V
\end{equation}
where the quantity in square brackets is our $A$ matrix, $T(E)$ is the $x$ array, and $V$ provide the $b$ arrays. 
There is an implied imaginary part in $(E-H_0)^{-1}$ where the limit is taken that the imaginary part goes to zero,
\begin{equation}
    \lim_{\eta\to0}\frac{1}{E-H_0+i\eta} = \frac{\mathrm{p.v.}}{E-H_0} - i\pi\delta(E-H_0),
\end{equation}
where $\mathrm{p.v.}$ denotes taking the Cauchy principal value.

\subsubsection{Interaction Potential}
Our interaction potential is the Coulomb interaction rather than the approximate dipole interaction.
The dipole interaction simplifies the atom-plasma interaction to be that of the inner product of the atomic dipole moment and the plasma electric field.
This approximation gives rise to the name ``Stark Broadening".
The full Coulomb potential that describes the interaction of two electrons can be expanded in terms of spherical harmonics,
\begin{equation}
    \frac{1}{|r_1-r_2|} = \sum_{kq}\frac{r_<^k}{r^{k+1}_>}\frac{4\pi}{2k+1}Y_{kq}^*(\hat{r}_1)Y_{lm}(\hat{r}_2).
\end{equation}
The total interaction matrix, $V$, for an $N$-electron atom that includes exchange is given by \citep{Fursa95}
\begin{equation}
    V = \sum_{i}V_{0i} + V_{0}^{(n)} + (E-H)\left[\sum_{i=1}^NP_{0i}\right]
    \label{Vmat}
\end{equation}
where the plasma/projectile electron is given coordinate $0$, while the set of $N$ radiator/target electrons are indices greater than zero.
The first two terms of Eq. \eqref{Vmat} are the direct matrix elements between the electrons and the nucleus and the remaining term is the exchange interaction.
In the exchange matrix element, $H$ is the Hamiltonian of the atom+projectile system and therefore only includes one electron in the broadening.
The explicit evaluation of exchange for one-electron atoms is explicitly given by \citet{GomezLyA}.
The extension for scattering (including exchange) for an $N$-electron system is given in \citet{Gomez26Coll}.

Due to the low density of these plasmas and magnetic field broadening, the exact details of multi-electron scattering will be less important.
In addition to the magnetic field broadening, the natural width will also significantly contribute to the Lorentzian width.

\subsection{Angular Coefficients for the Thermal Average with Fine Structure}

Pressure broadening is (generally speaking) an isotropic phenomenon. Therefore, certain symmetry properties can be exploited to simplify calculations.
Further exploitation of the type of radiation can also further simplify the problem and reduce the amount of work needed to calculate the broadening.
\citet{BenReuven66_2} simplified the broadening to (suppressing the frequency dependence and $k$ integrals)
\begin{widetext}
\begin{multline}
\mat{ab}{\EB}{a'b'} = n_e\sum_{k\ell}\rho(k)\Bigg[\sum_{\JT_i}\frac{[\JT_i]}{[J_a]}\delta_{bb'}\delta_{J_a,J_{a'}}\mat{a\ell}{T^{(\JT_i)}}{a'\ell} 
- \sum_{\JT_f}\frac{[\JT_f]}{[J_b]}\delta_{aa'}\delta_{J_b,J_{b'}}\mat{b\ell}{T^{*(\JT_f)}}{b'\ell}\\
+ 2\pi i\sum_{k'\ell'}\sum_{\JT_i\JT_f}(-1)^{\ell-\ell'+J_a-J_{a'}}[\JT_i,\JT_f]^{1/2}\sixJ{\JT_i}{\JT_f}{K}{J_b}{J_a}{\ell}\sixJ{\JT_i}{\JT_f}{K}{J_{b'}}{J_{a'}}{\ell'}\mat{a\ell}{T^{(\JT_i)}}{a'\ell'}\mat{b\ell}{T^{*(\JT_f)}}{b'\ell'},
\label{BenReuven}
\end{multline}
\end{widetext}
where $K$ is the multipole order of the radiation, for dipole radiation, $K=1$.
In this expression $ab$/$a'b'$ are the atomic state labels, $\ell$ is the partial wave of the perturbing electron, the $\JT$ is the total angular momentum of the combined atom+projectile system.
In our previous work, \citet{GomezLyA}, we did not couple the radiator to the spin and orbital angular momenta, so $L_a$ and $L_b$ was used in place of $J_a$ and $J_b$ and $\LT$ in place of $\JT$. 
In the de-coupled representation, there is an additional summation over spin with the coefficient $\delta_{S_a,S_b}\delta_{S_{a'},S_{b'}}[\ST]/[S_a,S_{a'}]^{1/2}$.

In the current work, however, we are interested in the broadening of each fine-structure component.
We therefore need to transform the $T$-matrices into a $jj$ representation and perform the trace in this representation.
The work of \citet{GomezLyA} did not previously account for fine structure in the interference term.

If separate fine-structure components are desired, then the upper and lower state broadening terms can be calculated directly from the $LS$ $T$-matrix \citep{Griem74}, i.e.
\begin{multline}
    \sum_{\JT_i}\frac{[\JT_i]}{[J_a]}\delta_{bb'}\delta_{J_a,J_{a'}}\mat{al}{T^{(\JT_i)}}{a'l} \\
    = \sum_{L_iS}\frac{[\LT_i,\ST_i]}{[L_a,S_a]}\delta_{bb'}\delta_{L_a,L_{a'}}\delta_{S_a,S_{a'}}\mat{al}{T^{(\LT_i,\ST_i)}}{a'l},
\end{multline}
and likewise for the lower state.
For the interference term, it's much more complicated, where the $T$-matrix in the $LS$ representation needs to be transformed into the $jj$ representation, via
\begin{multline}
    \mat{a (L_aS_a)J_a; (\ell s)j_p}{T^{\JT}}{a' (L_{a'}S_{a'})J_a; (\ell' s)j_p'} \\
    = \sum_{\LT \ST}T_{\LT\ST,jj}\mat{al_p}{T^{\LT \ST}}{a'l_p}T_{\LT\ST,jj}
\end{multline}
where the transformation matrix is given by
\begin{equation}
    T_{\LT \ST,jj} = [\LT,\ST,j_a,j_{p}]^{1/2}\nineJ{L_a}{\ell}{\LT}{S_{a}}{s}{\ST}{J_a}{j}{\JT}.
\end{equation}
Once the transformation to the $jj$ $T$-matrix, then the interference term of Eq. \eqref{BenReuven} can be used directly, but with $l$ replaced with $j$.

We next want to discuss the projection of the various broadening terms into their explicit $m$ dependence.
The upper-state and lower-state broadening terms do not require any modification, as these terms are diagonal in $M$.
For the interference term, then 
\begin{equation}
    (-1)^{j_{a}-m_a + j_{a'}-m_{a'}}[K]\threeJ{J_a}{K}{j_b}{-M_a}{Q}{M_b}\threeJ{J_{a'}}{K}{J_{b'}}{-M_{a'}}{Q}{M_{b'}}
\end{equation}
needs to be multiplied to the interference term to project the direction, which is needed in both ion and magnetic field perturbations.

\bibliographystyle{aasjournalv7}
\bibliography{bib}

\end{document}